\documentstyle[11pt,aaspp4,epsf,rotate]{article}

\def\psim{\lower.5ex\hbox{$\; \buildrel \propto \over \sim \;$}}

\begin{document}

\title{High Energy Gamma Rays from Ultrahigh Energy \\
Cosmic Ray Protons in Gamma Ray Bursts }

\author{Markus B\"ottcher\altaffilmark{1,2} \& Charles D.
Dermer\altaffilmark{2}}

\altaffiltext{1}{Department of Space Physics and Astronomy, Rice
University, 6100 Main Street, Houston, TX  77005-1892}
\altaffiltext{2}{E. O. Hulburt Center for Space Research, Code 7653,
       Naval Research Laboratory, Washington, DC 20375-5352}

\begin{abstract}

It has recently been proposed that ultrahigh energy ($\gtrsim 10^{19}$~eV)
cosmic rays (UHECRs) are accelerated by the blast waves associated with
GRBs.  We calculate the observed synchrotron spectrum from 
protons and energetic leptons formed in the cascades initiated by photopion
production, taking into account $\gamma\gamma$ attenuation at the source. 
Normalizing to the emission characteristics of GRB~970508, we predict 
$\sim 10$~MeV - 100~GeV fluxes at a level which may have been observed 
with EGRET from bright GRBs, and could be detected with the proposed 
GLAST experiment or with ground-based air \v Cerenkov telescopes having 
thresholds $\lesssim $ several hundred GeV. The temporal decay of the
UHECR-induced high-energy $\gamma$-ray afterglows is significantly
slower than that of the lower-energy burst and associated synchrotron 
self-Compton (SSC) 
radiation, which provides a direct way to test the hadronic 
origin of a high-energy GRB afterglow. Besides testing the UHECR origin 
hypothesis, the short wavelength emission and afterglows can be 
used to probe the level of the diffuse intergalactic infrared 
radiation field or constrain redshifts of GRB sources.

\end{abstract}


\section{Introduction}

The cosmological origin of GRBs has very likely been confirmed as a result of
observations of GRBs with the Beppo-SAX mission.  The good imaging of X-ray
emission from GRBs has led to optical counterpart identification of GRB 970228
(van Paradijs et al.\ \markcite{vanParadijs97}1997) and GRB 970508 (Djorgovski
et al.\ \markcite{Djorgovski97}1997).  In the case of GRB 970508,
observations of absorption lines in the spectrum of the optical 
counterpart place a lower limit on its redshift of $z = 0.835$ 
(Metzger et al.\ \markcite{Metzger97}1997).  

The blast wave model for GRBs has met with considerable success in
explaining the time dependence of the X-ray and optical afterglows (e.g.,
M\'esz\'aros, Rees, \& Wijers \markcite{MRW}1997; Vietri
\markcite{Vietri97a}1997a; Waxman \markcite{Waxman97}1997).  The release of
$\approx 10^{52}$ ergs of energy in a small volume results in the formation of a
relativistically expanding pair fireball which transforms most of the explosion
energy into the kinetic energy of baryons in a relativistic blast wave.
Reconversion of the kinetic energy into radiation occurs due to
collisions between different shells ejected from the central source 
or when the blast wave decelerates as it sweeps up matter from 
the external medium.

Shocks formed by these processes can accelerate protons to very high 
energies. If a significant fraction of the observed GRB power is 
transformed into UHECRs, then the measured energy density of these particles
is in rough agreement with the power
produced by GRBs, taking  into account that UHECRs travel $\lesssim 100$ Mpc 
before losing a 
significant fraction of their energy due to photopion production with 
the cosmic microwave background radiation (Zatsepin \& Kuzmin 
\markcite{Zatsepin66}1966; \markcite{Greisen66}Greisen 1966). On 
the basis of such arguments, Waxman
\markcite{Waxman95a}(1995a,\markcite{Waxman95b}b) and Vietri 
\markcite{Vietri95}(1995) proposed that UHECRs are accelerated by GRBs.   
Associated $\sim 10^{14}$ eV neutrino production (Waxman \& Bahcall 
\markcite{WB97}1997) and GeV photon production (Vietri 
\markcite{Vietri97b}1997b) due to UHECR acceleration during the 
prompt $\gamma$-ray emission phase of the GRB have been recently predicted.  
Detection of neutrinos  in temporal and spatial association with GRBs 
would confirm this cosmic ray origin theory, but will require larger
neutrino detectors than presently exist.  

In this {\it Letter}, we calculate both the prompt and delayed high 
energy $\gamma$-ray emission due to UHECRs accelerated in GRBs. We 
calculate the emergent synchrotron radiation from protons, from 
positrons produced in the decay of $\pi^+$, and from pairs produced 
by $\gamma\gamma$ interactions. The observed emission characteristics 
of GRB~970508 are used to set parameter values for the blast wave 
model of GRBs (see M\'esz\'aros \& Rees \markcite{MR93}1993;
M\'esz\'aros, Rees \& Papathanassiou \markcite{MRP94}1994; Dermer \& 
Chiang \markcite{Dermer98}1998). 

\section{Evolution of GRB Blast Wave}

After reaching a maximum speed defined by the baryon content and
total energy of the explosion, the Lorentz factor $\Gamma$ of a 
relativistic blast wave evolves by sweeping up matter from its 
surrounding medium (Blandford \& McKee \markcite{Blandford76}1976; 
M\'esz\'aros \& Rees \markcite{MR93}1993). For a blast wave of total
energy $E = 10^{52} E_{52}$~erg, which expands with initial bulk
Lorentz factor $\Gamma_0 = 300\, \Gamma_{300}$ into a medium of
uniform density $n_0$~cm$^{-3}$, the coasting phase ends and the 
deceleration phase begins when the blast wave has a characteristic 
radius given  by

\begin{equation}
x_0 = 2.6 \cdot 10^{16} \left( {E_{52} \over n_0 \, \Gamma_{300}^2}
\right)^{1/3} \> {\rm cm}
\end{equation}
(Rees \& M\'esz\'aros \markcite{Rees92}1992). For
$x > x_0$, the bulk Lorentz factor follows the power-law behavior 
$\Gamma (x) = \Gamma_0 \, (x / x_0)^{-g}$ where, for a uniform 
density medium, $g = 3/2$ in the non-radiative regime when 
transformation of the internal energy of swept-up particles into 
radiation is inefficient, and $g = 3$ in the radiative regime where the 
internal energy is efficiently radiated.

According to the scenario of UHECR acceleration by GRB blast waves 
(Vietri \markcite{Vietri95}1995; Waxman 
\markcite{Waxman95a}1995a,\markcite{Waxman95b}b), 
we assume that a fraction
$\xi$ of the energy  in nonthermal protons in the blast wave region is 
transformed into a 
power-law distribution $N_{\rm cr} (\gamma_{\rm cr};x) = K_{\rm cr}(x) 
\gamma_{\rm cr}^{-s}$ of ultrarelativistic cosmic ray protons with 
Lorentz factors $\Gamma (x) \le \gamma_{\rm cr} \le \gamma_{\rm cr,max}$ 
in the fluid frame comoving with a small element of the blast wave 
region. Here

\begin{equation}
K_{\rm cr}(x) = {1 \over m_p c^2} \, {\xi \, E_p (x) \, (2 - s) \over 
\gamma_{\rm cr,max}^{2 - s} - \Gamma(x)^{2 - s}}\; ,
\end{equation}
where $E_p(x)$ is the energy in non-thermal protons swept up from the
external medium, and the maximum cosmic-ray Lorentz factor is 

\begin{equation}
\gamma_{\rm cr,max} \simeq 10^{10} \, E_{52}^{1/3} \, \Gamma^{-2/3}
n_0^{1/6} \zeta^{1/2} r^{1/2}
\end{equation}
(Vietri \markcite{Vietri95}1995, \markcite{Vietri97b}1997b).  The term $\zeta$ 
in this equation is the
equipartition factor for the magnetic field, given through the expression
\begin{equation}
H({\rm G}) = (8\pi r m_pc^2n_0)^{1/2}\Gamma \zeta^{1/2} \cong 0.6\; \Gamma
[(r/10) \> n_0  \> \zeta]^{1/2}\;,
\end{equation}
and $r$ is the compression ratio which can exceed 4 for relativistic 
shocks or in shocks where nonlinear feedback of the nonthermal
particle pressure affects the shock structure (e.g., Ellison, Jones, \& 
Reynolds \markcite{Ellison90}1990).  We let $\xi = 1$, $r=10$, and $\zeta = 1$ 
in our 
calculations. The choice of $\zeta \sim 1$ is needed for 
Fermi acceleration to be efficient enough to produce cosmic-ray protons 
of $E_p \sim 10^{20}$~eV (Vietri \markcite{Vietri95}1995).

Finally, we have to relate the measured flux at a given time to the 
comoving  differential photon density $n_{\rm ph} (\epsilon)$, where 
$\epsilon = E_{\gamma} / (m_e c^2)$ is the dimensionless photon energy. 
The observed radiation is a convolution of the contributions from 
regions of the blast wave emitting at different times and moving with 
different Lorentz factors and directions with respect to the observer 
(Rees \markcite{Rees1967}1967). The completely self-consistent 
transformation to the observer's frame can therefore in general 
only be done numerically, taking into account the evolution of 
the intrinsic radiation spectrum. However, due to the very strong 
Doppler enhancement along the direction of motion in a relativistic 
blast wave, only a small fraction of the blast wave covering a solid 
angle of order $1/\Gamma^2$ contributes significantly to the observed 
radiation, within which the difference in light travel time to the 
observer is negligible. Within this approximation, an observed power-law 
spectrum given by $S_{\rm obs} (\epsilon_{\rm obs}) = S_0 \, 
\epsilon_{\rm obs}^{-\alpha}$ between $\epsilon_{\rm obs, 1} \le 
\epsilon_{\rm obs} \le \epsilon_{\rm obs, 2}$ is related to  the 
spectral photon density in the comoving frame, given by $n_{\rm ph} 
(\epsilon) = n_{\rm ph}^0 \, \epsilon^{- (1 + \alpha)}$ between 
$\epsilon_1 = (1 + z) \, \epsilon_{\rm obs, 1} / (2\Gamma) \le 
\epsilon \le (1 + z) \, \epsilon_{\rm obs, 2} / (2\Gamma) = 
\epsilon_2$, through the expression

\begin{equation}
n_{\rm ph}^0 \cong {4 \, d_L^2 \, S_0 \, (1 + z)^{\alpha - 1} \over
c \, x^2 \, m_e c^2 \, (2 \Gamma)^{1 + \alpha}}.
\end{equation}
Here, $d_L$ is the luminosity distance. 

\section{Photopion Production, Proton Synchrotron Radiation
and Pair Cascades}

We calculate the $\gamma$-ray and positron production spectra from 
decaying pions produced by UHECRs interacting with photons (see, e.g., 
Stecker \markcite{Stecker79}1979). The UHECR energy-loss rate through 
photopair production is much smaller than that of photopion production 
for the proton energies and photon spectra considered here, and will 
be neglected. The differential cross section for photopion production is 
approximated by $d\sigma / d \epsilon' \, \approx \, \sigma_0 \delta 
(\epsilon' - E_{\Delta}/m_ec^2)$, where $\sigma_0 = 2 \cdot 10^{-28}$~cm$^2$, 
$\epsilon'\cong \gamma_{\rm cr}\epsilon(1-\mu)$ is the photon 
energy  in the proton rest frame, and $E_{\Delta} \cong  330$~MeV. 
Noting that pions are produced primarily near threshold in the protons' 
rest frame (i. e., $\gamma_{\pi} \approx \gamma_{\rm cr}$), we obtain 
the pion production rate

$$
\dot N_{\pi} (\gamma_{\pi}) = {c \over 2} \int\limits_1^{\infty} d\gamma_{\rm 
cr}
\> N_p (\gamma_{\rm cr}) \int\limits_{-1}^1 d\mu \int\limits_0^{\infty}
d\epsilon \> n_{\rm ph} (\epsilon) \, (1 - \beta_p \mu) \, {d^2 \sigma
\over d\mu \, d\gamma_{\pi}}
$$
$$ \approx \sigma_0 n_{\rm ph}^0 \, { c \, K_{\rm cr}
\, E_{\Delta} \over 2 \, m_e c^2 \, (\alpha + 1)} \, \gamma_{\pi}^{-(1 + s)}
\, \left\lbrace  \left[ \max \left( \epsilon_1, {E_{\Delta} \over 2 \, 
\gamma_{\pi} \, m_e c^2} \right) \right]^{-(\alpha + 1)} - 
\epsilon_2^{-(\alpha + 1)} \right\rbrace \> \cdot
$$
\begin{equation}
\cdot \> \Theta\left\lbrace \gamma_{\pi}; \, \max\left[\Gamma(x), 
{E_{\Delta} \over 2 \, m_e c^2 \, \epsilon_2} \right], \, 
\gamma_{\rm cr,max} \right\rbrace.
\end{equation}
The generalized Heaviside function $\Theta$ is defined as $\Theta
(x; a, b) = 1$ if $a \le x \le b$, and $0$ otherwise. The photon spectrum
from decaying neutral pions is then

$$
\dot N_{\gamma}^{ \pi^0 \to 2\gamma} (\epsilon) = \sigma_0 \, n_{\rm ph}^0
\, { c \, K_{\rm cr} E_{\Delta} \over 2 \, \, m_{\pi} c^2 \, (\alpha + 1)} 
\> \cdot
$$
\begin{equation}
\cdot \!\!\!\! \int\limits_{\max\left[ \epsilon m_e / m_{\pi}, 
\, \Gamma(x), \, E_{\Delta} / (2 \, \epsilon_2 \, m_e c^2)
\right]}^{\gamma_{\rm cr,max}} \!\!\!\!\! d \gamma_{\pi} \, 
\gamma_{\pi}^{-(2 + s)} \, \left\lbrace \left[ \max \left( 
\epsilon_1, {E_{\Delta} \, \over 2 \, \gamma_{\pi} \, m_e c^2} \right) 
\right]^{-(\alpha + 1)} - \epsilon_2^{-(\alpha + 1)} \right\rbrace.
\end{equation}

For simplicity, we assume that charged pions are produced at the
same rate (Eq. [6]) as $\pi^0$s and, after decaying into one 
positron and three neutrinos, on average $1/4$ of the energy of 
the pion is carried by the positron (instead of $1/2$ carried by 
each of the two photons in the case of $\pi^0$ decay). This yields
$\dot N_{e^+} (\gamma_+) \approx \dot N_{\gamma}^{\pi^0 \to 2\gamma}
\left( 2 \, \gamma_+ \right)$ for the first generation of relativistic 
positrons injected into the shock region. We take into account energy 
losses due to synchrotron radiation, but neglect escape, pair 
annihilation, and Compton losses.  Escape is unlikely 
given the small Larmor radius, and pair annihilation is only 
important for low-energy positrons which radiate inefficiently.  
Compton losses are found to be much less important than 
synchrotron losses for the  parameters used here. Integration of 
the continuity equation  yields  the ``steady-state'' distribution 
of the first generation of positrons.

The proton synchrotron spectrum is calculated using a $\delta$-function
approximation for the emission of a single proton, giving

\begin{equation}
\dot N_{\gamma} (\epsilon) = {c \sigma_{\rm T} \, H_c^2 \over (27 \, 
\pi \, m_e c^2)} \, \left( {\epsilon_{H, p} \over \epsilon} \right)^{1/2} 
N_{\rm cr} \left( \sqrt{\epsilon \over \epsilon_{H, p}} \right),
\end{equation}
where $\epsilon_{H, p} = (3/2) (m_e / m_p) \, H / H_{c}$, and $H_{c} = 4.414 
\cdot 10^{13}$~G.

 We compute the opacity $\tau_{\gamma\gamma}$ using the full pair production 
cross section and assuming that the shell width $\delta x_{\rm sh} 
= x/\Gamma$ (M\'esz\'aros, Laguna, \& Rees \markcite{MLR93}1993). The shock 
region is optically
thick to $\gamma\gamma$ pair production on the soft burst and afterglow 
radiation for photons 
with $E_{\gamma} \gtrsim 100$~GeV. This differs from the results
of Waxman \& Bahcall (1997) and Vietri (1997b), who find the region
to be optically thick for photons above $\sim 100$~MeV and $\sim 1$~TeV, 
respectively. This
difference is largely due to the smaller size scale
used by Waxman \& Bahcall and the different nominal values of the Lorentz 
factors ($\Gamma_{300} =
1/3$, 1, and 3 in Waxman \& Bahcall, this study, and Vietri, 
respectively). The $\gamma\gamma$ absorbed portion of the high-energy 
photons initiates a pair cascade. Since most of the secondary 
pairs of the electromagnetic cascade are injected at high energies, 
the steady-state pair distribution of secondary particles can 
be approximated by a power-law with index 2 over a wide 
range of electron/positron energies. This yields a 
synchrotron spectrum of spectral index $\alpha_{\rm syn} = 1/2$. 
We approximate the radiative output of the cascade by a 
$\gamma\gamma$-absorbed power-law spectrum of energy index 
$1/2$ containing the energy of the absorbed high-energy
photons from $\pi^0$ decay and synchrotron radiation of
the first generation of positrons from $\pi^+$ decay. 

SSC radiation from the primary electrons may also play an important
role. The numerical simulations described in Chiang \& Dermer
(\markcite{chiang98}1998) were used to determine the expected level
and energy range of the SSC component.  Note that the SSC component is
significantly weaker than estimated using simple Thomson-limit  arguments 
because Klein-Nishina effects are very important in this  case.

\section{Numerical results}

For the purpose of illustration, we compute the expected high-energy prompt
radiation and afterglow from GRB~970508. We approximate its observed spectrum
during the burst phase by a broken power-law with energy index $\alpha_1 = 0$ 
below $E_{\rm break} \approx 45$~keV and $\alpha_2 = 1$ above the break, 
fitting to the Beppo-SAX fluxes measured in the 2-26 keV and 40-700 keV 
bands during the first 25 s of the GRB (Piro et al. \markcite{Piro98}1998). 
The spectrum is assumed to extend from $E_{\rm obs, 1} = 1$~eV, 
corresponding to the self-absorption frequency early in the 
GRB, to $E_{\rm obs, 2} = 10$~MeV (the calculations do not
depend sensitively on the upper energy). We assume $E = 10^{52}$~erg, 
$z = 0.835$, $\Gamma_0 = 300$. Eq. (3) implies that the cosmic-ray 
spectrum extends up to $\gamma_{\rm cr,max} = 7 \cdot 10^8$, and 
we assume that it has a spectral index of $s = 2$. Fig. 1 shows 
the different  components of the resulting broad-band spectrum 
at the point $x_0$,  and the pair production opacity 
$\tau_{\gamma\gamma}$ at this point.

Examing synchrotron losses in the evolving magnetic field (eq. [4]), 
one finds that only particles with $\gamma \lesssim 5(m/m_e)^3/[ (r/10)\zeta
n_0^{2/3}\Gamma_{300}^{1/3} E_{52}^{1/3}$] cool efficiently through 
synchrotron proceses. Thus synchrotron  cooling is very inefficient 
for protons, though electrons cool efficiently if $\zeta \approx 1$. 
The radiative output of protons is dominated by synchrotron radiation 
rather than photo-pion production, although the efficiency of the 
latter process can be increased in a colliding shell scenario 
(e.g., Waxman \& Bahcall \markcite{WB97}1997).  

Approximating the observed evolution of the optical to soft
$\gamma$-ray spectrum of the GRB and the afterglow from  GRB~970508, 
we assume that the flux remains constant for $t_0 = 30$~s in the 
observer's frame and then decays as $S (t) = S_0 \,
(t/t_0)^{-1.07}$. The break energy  $E_{\rm break}$ and the upper
cut-off $E_2$ are assumed to decrease with the same temporal 
behavior. Because cooling is inefficient, we expect that the 
evolution of the blast wave is more closely described by the 
non-radiative limit. Fig.~2 illustrates the time dependence 
of the high-energy $\gamma$-ray afterglow, compared to the 
evolution of the low-energy burst and the SSC radiation 
in this case. The high-energy afterglow decays much more slowly 
than the optical--to--X-ray afterglow and the SSC component.  
In the radiative case, the proton synchrotron afterglow radiation 
decays much faster than in the non-radiative case, but still 
considerably less rapidly than the optical--to--X-ray 
afterglow.

The temporal behavior of the uncooled proton synchrotron flux in 
the evolving field (4) goes as $t_{\rm obs}^{-\chi}$  where 
$\chi_{\rm uncooled} =  [4g + (7\eta/4) - 3]/(1+2g)$ for $s=2$  
and an external medium $n_{\rm ext}\propto x^{-\eta}$ (see Dermer 
\& Chiang \markcite{Dermer98}1998, eqs. [50], [51]). In the cooled 
regime, characteristic of the electon synchrotron  
radiation, we find $\chi_{\rm cooled} = (4g + \eta
-2)/(1+2g)$.   In the non-radiative $g=3/2$ regime with $\eta = 0$, 
we find that $\chi_{\rm uncooled} = 3/4$ and $\chi_{\rm cooled} = 1$.  
The effects of the cascade cause the high-energy radiation to decay 
even more slowly than estimated by these analytic expressions.

The level of the proton synchrotron flux is not strongly dependent 
on the endpoints of the low-energy spectrum because this component 
is produced mainly at energies where intrinsic $\gamma\gamma$ 
attenuation is unimportant.  The photopion component is, however, 
reduced when the high-energy cutoff of the burst emission is decreased. 
Due to the smaller $\gamma\gamma$ optical depth, a smaller portion of 
the high-energy radiation resulting from $\pi^0$ decay and synchrotron 
emission of first-generation positrons is absorbed and transferred to 
the cascade. Different values of $\Gamma_0$ significantly change the 
energy range and amplitude of the emergent proton synchrotron and 
photopion emission, but only during the early portion of the GRB.  
At comparable times later in the GRB, the flux and energy ranges 
are similar.

\section{Discussion}

The high energy $\gamma$ radiation from
proton synchrotron and the photopion-induced pair cascade decays 
more slowly than the low energy afterglow and the SSC radiation, 
because cooling is inefficient for protons whereas
electrons cool efficiently. Therefore, synchrotron radiation from 
protons can be distinguished from direct electron synchrotron or 
SSC emission by comparing the temporal decay rates in the afterglow 
phase, even though the SSC radiation is produced at roughly the same flux 
levels as the proton synchrotron radiation early in the GRB.  A monotonically 
increasing ratio of
the high-energy $\gamma$-ray  flux to the optical or soft X-ray flux in the GRB 
afterglow would 
provide strong evidence for proton synchrotron radiation and 
UHECR acceleration in GRBs. 

The flux sensitivities of EGRET and GLAST (pointed mode) 
for an $\alpha = 1$ spectrum are $ \approx 8 \cdot 10^{-8}$ ph 
($> 100$ MeV) cm$^{-2}$ s$^{-1}$ and $ \approx 2 \cdot 10^{-9}$ 
ph ($> 100$ MeV) cm$^{-2}$ s$^{-1}$, respectively (Kurfess et al.
\markcite{Kurfess97}1997; GLAST \markcite{Glast97}1997). In the 
inset to Fig.~2, we compare the fluence sensitivities 
of EGRET, GLAST and Whipple to the fluence predicted 
by our calculations of GRB~970508.  The calculated 
$> 100$ MeV burst fluence is an order of magnitude below the EGRET
sensitivity level, but could be detected in brighter GRBs. GLAST,
when operated in the pointed mode, would be able to detect the
$> 100$~MeV $\gamma$-ray afterglow at the predicted level. The
emission from UHECRs accelerated in GRBs may already have been 
observed in six bright BATSE bursts that were detected with EGRET 
(Dingus \markcite{Dingus95}1995; Catelli, Dingus, \& Schneid
\markcite{Catelli96}1996; Hurley et al.\ \markcite{Hurley94}1994)
and display weak evidence for a hard tail at $> 100$ MeV and GeV 
energies. From Fig.~2, we note that the best opportunity for 
detection of $\sim 100$ MeV emission with EGRET or GLAST is within 
the first several minutes of the GRB whereas a larger fraction of 
the high-energy gamma-ray emission is produced on time scales of 
hours, which is in accord with the detection of an 18 GeV photon from 
GRB~940217 90 minutes after the main part of the GRB (Hurley et al. 
\markcite{Hurley94}1994).

Fig.~2 indicates weak delayed $> 100$ GeV-TeV emission, which is below 
the thresholds of existing \v Cerenkov telescopes.  This emission would not, 
in any case, be detectable from a GRB located at a redshift $z \gtrsim 0.5$ 
(see Stecker \& de Jager \markcite{Stecker97}1997; Salamon \& Stecker 
\markcite{Salamon98}1998) due to absorption by the intergalactic 
infrared radiation field, but could be detected from nearby bright 
GRBs. The delayed afterglows in the 10-100 GeV range from such GRBs 
could provide a way to measure the level of the infrared background 
radiation and to yield upper limits on redshifts, but will require 
high sensitivity ground-based \v Cerenkov telescopes with low energy 
thresholds. If redshifts are independently measured, cutoffs in the 
observed high energy radiation spectrum can be used to infer the 
level of the infrared background.

The photopion process could be more important during the early 
phase of the GRB if the size scale of the system is determined by 
collisions between shells (Waxman \& Bahcall \markcite{WB97}1997).  
In this case, we would expect a considerably larger $> 10$ MeV 
flux due to the photopion/$\gamma\gamma$ cascade  during the 
primary $\gamma$-ray emission phase. Flux levels of high-energy
$\gamma$ radiation could therefore constrain the expected level of 
neutrino production from UHECRs and discriminate between the colliding 
shell and external shock scenario of GRBs (Vietri \markcite{Vietri97b}1997b).

\acknowledgements
We thank J. Chiang for useful discussions and the anonymous referee for 
a detailed and constructive report. MB acknowledges support by 
the German Academic Exchange Service (DAAD). The work of CD is 
supported by the Office of Naval Research and the {\it Compton Gamma Ray 
Observatory}
Guest Investigator program.

\eject

\begin{figure}
\rotate[r]{
\epsfysize=12cm
\epsffile[300 50 600 500]{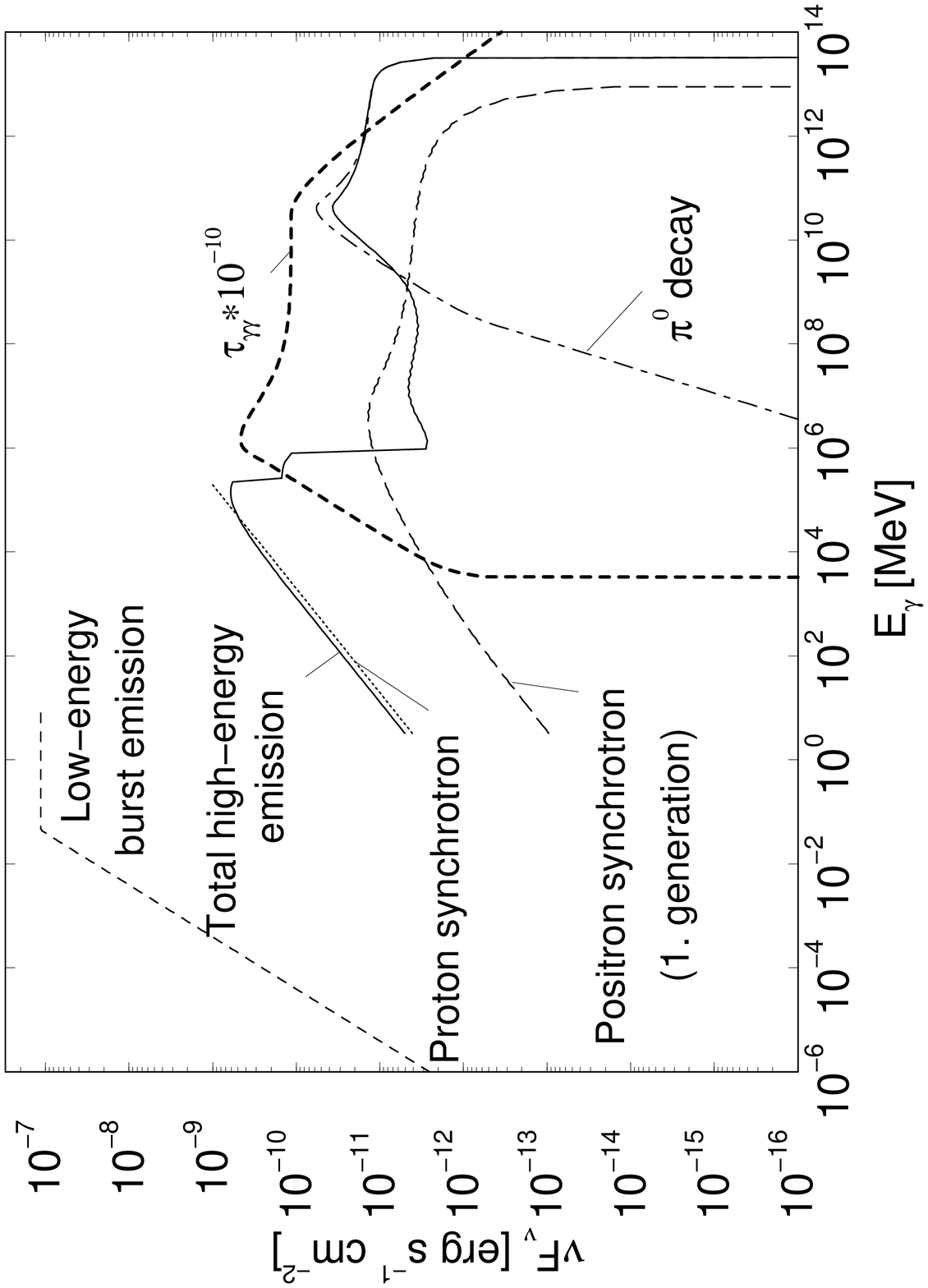}}
\caption[]{Broadband spectrum of a GRB blast wave at the point $x_0$
of transition to the asymptotic regime, for parameters derived from
the observations of GRB~970508. Dashed: optical to soft
$\gamma$-ray spectrum; dotted: proton synchrotron radiation; 
long dashed: synchrotron radiation from positrons produced 
in $\pi^+$ decay; dot-dashed: $\pi^0$ decay; solid: total
high-energy spectrum, including $\gamma\gamma$ absorption 
and synchrotron emission from the pair cascade; thick dashed: 
${\gamma\gamma}$ pair production opacity multiplied by $10^{-10}$.}
\end{figure}

\eject

\begin{figure}
\epsffile[0 350 750 800]{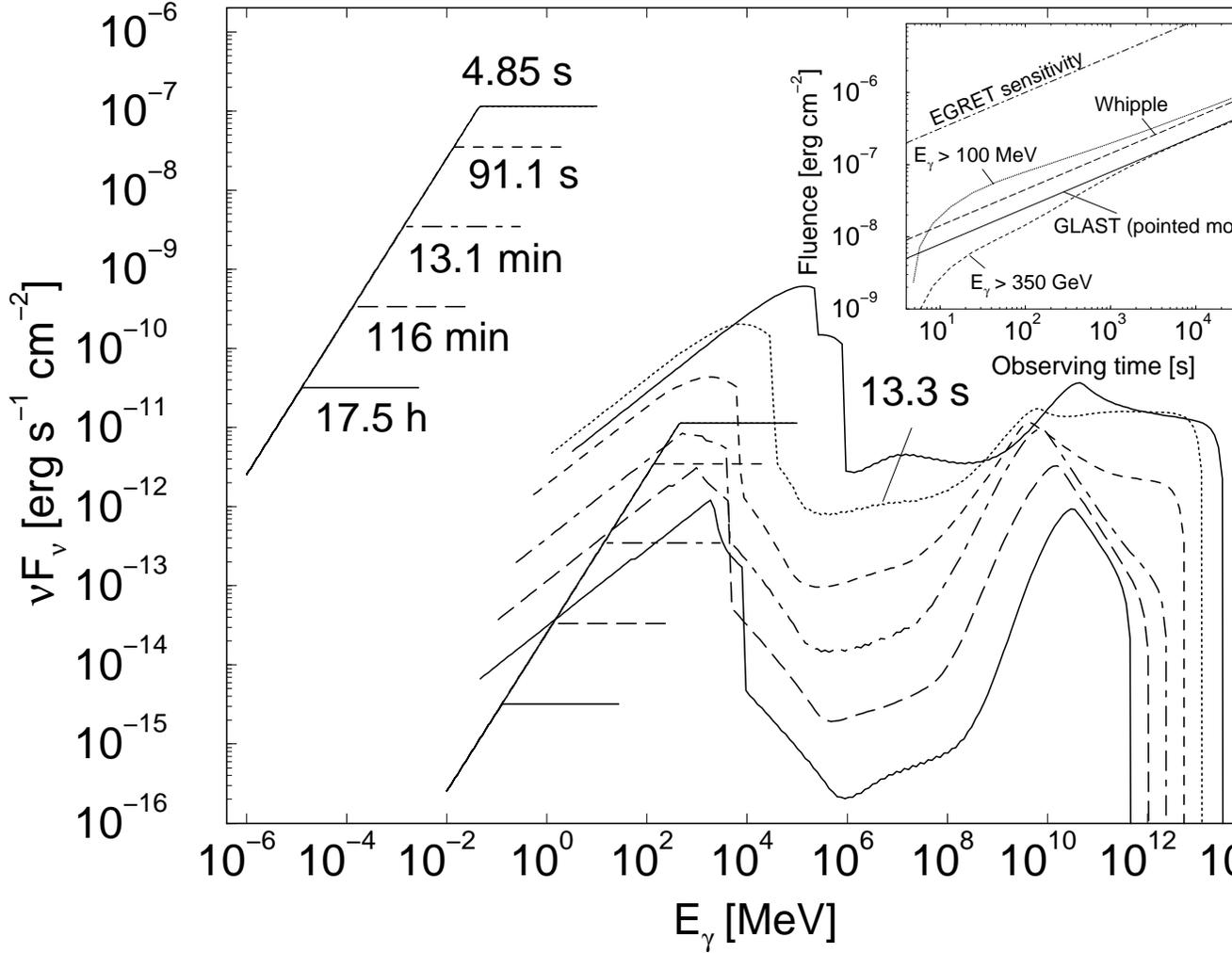}
\caption[]{Time history of the broadband burst and afterglow emission 
in the non-radiative regime at different observer times. The expected
SSC radiation associated with the low-energy burst emission is 
shown for comparison. The inset shows time- and 
energy-integrated fluxes for high energy emission from 
GRB~970508 for energies $> 100$ MeV (dotted curve) and $>350$ GeV 
(short-dashed curve). The dot-dashed, solid and long-dashed line 
are the fluence sensitivities for EGRET, GLAST in its pointed mode,
and Whipple, respectively.}
\end{figure}

\end{document}